\documentclass[fleqn,twoside]{article}
\usepackage{espcrc2}

\usepackage{graphicx}
\usepackage[figuresright]{rotating}

\newcommand{\AmS}{{\protect\the\textfont2
  A\kern-.1667em\lower.5ex\hbox{M}\kern-.125emS}}

\title{Observing Gravitational Radiation with QSO Proper Motions and the
SKA}

\author{Andrew H. Jaffe\address{Astrophysics, The Blackett Laboratory, Imperial College, London SW7 2AZ UK\\\texttt{a.jaffe@imperial.ac.uk}}}
       
\begin{document}

\begin{abstract}
  We discuss the ability of the SKA to observe QSO proper motions
  induced by long-wavelength gravitational radiation. We find that the
  SKA, configured for VLBI with multiple beams at high frequency (8
  GHz), is sensitive to a dimensionless characteristic strain of roughly
  $10^{-13}$, comparable to (and with very different errors than) other
  methods in the 1/yr frequency band such as pulsar timing.

\vspace{1pc}
\end{abstract}

\maketitle

\section{Introduction}

The history of astronomy has been that of our measurement of photons
from astrophysical objects and, more recently, the early universe (the
CMB). Analogously, the universe is thought to be suffused by
gravitational radiation --- a background from sources in the early
universe (e.g., inflation) and ``foregrounds'' from the gravitational
interaction of massive astrophysical objects, such as the massive black
holes in the centers of galaxies. This gravitational radiation perturbs
spacetime, altering path-lengths and lensing light rays, throughout the
universe. The small amplitude of the perturbations (strains of
$h\sim10^{-15}$) makes these effects very difficult to observe. In this
section, we examine the time-dependent ``gravitational lensing'' which
results in apparent proper motions of distant galaxies, potentially
observable by the SKA.

\section{GW-induced Proper motions}

The pattern of proper motions induced by a single gravity wave with
``+'' polarization, amplitude $h$, and frequency $\omega$ in the ${\hat
  z}$ direction is \cite{Pyne,KJ,Gwinn}
\begin{eqnarray}
  \label{eq:pattern}
  \mathbf{\mu}(\omega,h,{\hat z}) &=& \frac12 \omega h \sin\omega t \sin\theta ({\hat\theta}\cos2\phi-{\hat\phi}\sin2\phi)\; .\nonumber\\
\end{eqnarray}
The pattern from an arbitrary wave can be calculated by superposition of
this pattern at different amplitudes and directions. Obviously, the
particular pattern is not preserved by superposition, but the power
spectrum of the pattern is \cite{Gwinn}. We calculate the power spectrum
by decomposing the pattern into \textbf{vector spherical harmonics},
$\mathbf{Y}_{\ell m}^{A}({\hat x})$, ($A=E,B$) the generalization of the
usual scalar functions $Y_{\ell m}$ to vectors defined on the sphere. In
fact, these are closely analogous to the tensor spherical harmonics used
in the study of weak lensing and of the polarization of the CMB. Indeed
the goal -- to measure the power spectrum of the geometrical pattern --
is much the same in both cases. We decompose the total pattern into 
\begin{eqnarray}
  \label{eq:alm}
  \mathbf{\mu}(\omega)&=&\sum_{\ell,m,A} a_{\ell m}^A \mathbf{Y}_{\ell m}^A
  \nonumber\\
  &=& \omega h_c(\omega) \sum_{\ell,m,A} {\hat a}_{\ell m}^A \mathbf{Y}_{\ell m}^A
\end{eqnarray}
where the coefficients $a_{\ell m}^A$ are functions of frequency and the
${\hat a}_{\ell m}$ separates out the spatial and frequency
dependence. The quantity $h_c(\omega)$ is the characteristic strain at a
given frequency \cite{JaffeBacker}. We can finally define the power spectra
\begin{eqnarray}
\label{eq:MAl}
M_\ell^{A}(\omega)=\langle |a_{\ell m}^A|^2\rangle &=& \omega^2 h_c^2(\omega) \langle |{\hat a}_{\ell m}^A|^2\rangle \nonumber\\
&=& \omega^2 h_c^2(\omega){\hat M}^A_\ell
\end{eqnarray}
  
However, we do not directly observe the proper motion, $\mu$, itself, which is an angular velocity. Rather, we measure the position of objects at two (or more) epochs separated by time $\delta t$ -- the integral of the angular velocity. The power spectrum of this position difference is then an integral over the $\mu$ power spectrum. In analogy to Eq.~\ref{eq:alm}, we define $\theta_{\ell m}(\delta t)$ as the transform of the position difference pattern, with power spectrum 
\begin{equation}
\label{eq:MlA2}
\langle \left|\theta_{\ell m}(\delta t)^A\right|^2\rangle=
 2\int \frac{d\omega}{\omega^2}\; M_\ell^{A}(\omega) 
 \left[1-\cos(\omega\delta t)\right].
\end{equation}

In fact, Gwinn et al \cite{Gwinn} show that a full $5/6$ of the power in
gravitational radiation is at $\ell=2$ (the quadrupole); in practice
there is little need to measure the entire spectrum. However, the
\emph{time}-frequency ($\omega$) spectrum depends on the frequency
content of the gravitational radiation background:
\begin{equation}
  \label{eq:OmegaGW1}
  M_\ell^{A}(\omega) =\omega h_c^2(\omega) \left|{\hat a}_{\ell m}\right|^2
\end{equation}
and
\begin{equation}
  \label{eq:OmegaGW2}
  M_2^{E} + M_2^{B} = \frac{2\pi}{3}H_0^2 \Omega_\mathrm{GW}(\omega)
\end{equation}
where $\Omega_\mathrm{GW}(\omega)$ gives the contribution to the
cosmological critical density of the GW background per logarithmic
integral of frequency.

\section{Proper motion error analysis}

From the study of the CMB, we recall that the error at a particular
multipole is given by \cite{Knox95,HobMag96}
\begin{equation}
  \label{eq:error}
  (\delta M_\ell)^2 \simeq \frac{2}{(2\ell+1)f_\mathrm{sky}} \left(M_\ell + w^{-1}\right)^2
\end{equation}
where $w$ gives the weight (inverse variance) per solid angle and we
have assumed observations uniformly spread around a fraction
$f_\mathrm{sky}$ of the sky. This formula is just the usual one for the
variance of the square of a quantity with known variance:
$(2\ell+1)f_\mathrm{sky}$ is the number of modes at a given $\ell$.  The
$M_\ell$ term then gives the sample variance, and the $w^{-1}$ term the
noise variance (i.e., resulting from interferometric phase errors).

It remains, then, only to insert the expected signal for $M_\ell$ and
the noise characteristics of the SKA. In the band of frequencies probed
by the SKA, the dominant contribution is likely to be massive black
holes (MBHs) in close binaries at the centers of galaxies:
\begin{eqnarray}
\label{eq:MBHhc}
h_c(\omega)^2 &\sim& \left(10^{-15}\right)^2 \nonumber\\&& \times 
\left(\frac{\omega}{\mathrm{yr}^{-1}}\right)^{-4/3}\langle\left(\frac{\cal
    M}{10^7M_\odot}\right)^{5/3}\rangle  I_{h^2} 
 \; .\nonumber\\
\end{eqnarray}
where ${\cal M}^{5/3}=M_1M_2(M_1+M_2)^{-1/3}$ gives the so-called
``chirp mass'' of the system, angle brackets denote the average over the
massive black hole mass function, and the dimensionless factor $I_{h^2}$ gives an integral over the merger rate of black holes \cite{JaffeBacker}.

Beneath this foreground potentially lies the cosmological background
from an epoch of inflation \cite{Maggiore}:
\begin{eqnarray}
\label{eq:inflhc}
h_c^2(\omega) &\sim& \left(10^{-17}\right)^2 \nonumber\\ && \times
\left(\frac{H_*}{10^{-4}M_\mathrm{Pl}}\right)^2 \left(\frac{\omega}{\mathrm{yr}^{-1}}\right)^{n_T-2}10^{9n_T}\nonumber\\
\end{eqnarray}
where $H_*$ is the energy scale of inflation, $n_T$ is the Tensor spectral index from inflation (usually, $n_T\simeq0$ and $n_T<0$), and we have ignored the so-called running of the index; this formula gives a generous upper bound and thus any realistic background is well below any astrophysical foregrounds such as Eq.~\ref{eq:MBHhc}.

From Eqs.~\ref{eq:MAl}--\ref{eq:MlA2}, we see that we need integrals of the form
\begin{equation}
\int d\omega h_c^2(\omega)\left[1-\cos(\omega\delta t)\right]\; .
\end{equation}
For these power-law spectra, we of course just pick out the amplitude at $\omega\sim1/\delta t$. 

\section{Discussion}

For the proposed measurements with the SKA, we will likely only be
placing limits on the gravitational radiation background and so the
noise term will be dominant. The weight per solid angle is given by $w=N
\sigma^{-2}/(4\pi f_\mathrm{sky})$ for $N$ proper motion observations
with error $\sigma^2$ on each observation. From Ref.~\cite{FomRei}, we
can expect an rms position accuracy of $10\mu \mathrm{as} =
5\times10^{-11}\mathrm{rad}$ with the SKA used in VLBI mode at (say) 8
GHz. Without ``multibeams,'' the SKA will be able to measure the
positions of roughly $10^4$ point-like QSOs in about one month; with 100
multi-beams this increases to the roughly $10^6$ objects available on
the sky \cite{Rawlings}. This gives a total weight $w^{-1} =
(10^{-13})^2$ for the multibeam configuration and $f_\mathrm{sky}=1/2$.
This indicates that our sensitivity to gravitational radiation will be
at a level of roughly $h\sim10^{-13}$, competitive with other methods
such as pulsar timing \cite{JaffeBacker}. In more detail, we see that
this noise contribution will likely still be dominant for
$\omega\sim\mathrm{yr}^{-1}$ for the signals in
Eqs.~\ref{eq:MBHhc}--\ref{eq:inflhc}. Without multi-beaming, we of
course lose another factor of 10 in sensitivity to $h$, and the method
is less competitive. However, this calculation assumes that the position
errors are uncorrelated; in practice phase errors from the troposphere
will correlate position errors, and this can mimic the low angular
frequency pattern of the gravitational waves. We have also ignored the
loss of some low-frequency information to the ``geoedetic'' information
required for VLBI, as discussed in Ref.~\cite{Gwinn}. Nonetheless, these
are very different systematic problems than those to which pulsar timing
will be sensitive.

We thus propose to use the SKA configured for VLBI in multibeam mode at
8 GHz, observing a preselected population of $10^6$ pointlike AGN in two
epochs, separated by at least one year (more epochs and a longer time
difference both increase the sensitivity).

\end{document}